\documentclass[11pt]{article}
\usepackage{times}                                                                                                                                
\usepackage{soul}
\usepackage{url}
\usepackage[hidelinks]{hyperref}
\usepackage[utf8]{inputenc}
\usepackage[small]{caption}
\usepackage[final]{graphicx}
\usepackage{algorithm}
\usepackage{algorithmic}
\usepackage[switch]{lineno}

\usepackage{amsmath,amsfonts,amsthm,amssymb,color}
\usepackage{fancybox}
\usepackage{cleveref}
\usepackage{verbatim}
\usepackage{float}
\usepackage{booktabs}

\usepackage{siunitx}
\graphicspath{{Images/}}

\newenvironment{fminipage}%
{\begin{Sbox}\begin{minipage}}%
		{\end{minipage}\end{Sbox}\fbox{\TheSbox}}

\theoremstyle{definition}

\def\defeq{\stackrel{\mathrm{def}}{=}}

\newcommand\rewarddiv{\lambda_{\mathsf{div}}}
\newcommand\pencoauthor{\lambda_{\mathsf{co}}}
\newcommand\pencycle{\lambda_{\mathsf{cyc}}}

\crefname{equation}{equation}{equations}

\allowdisplaybreaks

\begin{document}

\title{A Unified Framework for Scalable and Robust Paper Assignment}



\author{
Michael Cui $^{1}$ \thanks{Equal contribution.} \quad
Chenxin Dai $^{2}$ \footnotemark[1] \quad
Yixuan Even Xu$^{1}$ \quad
Fei Fang$^{1}$ \\
\\
$^{1}$ Carnegie Mellon University \\
$^{2}$ Independent Researcher \\
\\
\texttt{\{mjcui, yixuanx, feif\}@andrew.cmu.edu} \\
\texttt{gtrhetr@gmail.com}
}

\maketitle

\begin{abstract}
Assigning papers to reviewers is a central challenge in the peer-review process of large academic conferences. Program chairs must balance competing objectives, including maximizing reviewer expertise, promoting diversity, and enhancing robustness to strategic manipulation, but it is challenging to do so at the modern conference scale.

Existing algorithmic paper assignment approaches either fail to address all of these goals simultaneously or suffer from poor scalability. To address the limitation, we propose Robust Assignment via Marginal Perturbation (RAMP), a unified framework for large-scale peer review. Our approach formulates a linearized perturbed-maximization objective with soft constraints that flexibly balance assignment quality, diversity, and robustness while maintaining runtime efficiency. We further introduce an attribute-aware sampling procedure that converts fractional solutions into integral assignments and improves the diversity and robustness of the final assignment. On datasets with over $20,000$ papers and $20,000$ reviewers, RAMP runs in under $20$ minutes, demonstrating its suitability for real-world deployment.

\end{abstract}

\section{Introduction}\label{sec:intro}

Most scientific venues rely on peer review to ensure the quality and integrity of published research \cite{Protasiewicz20143062}. Traditionally, paper assignment was handled manually or via simple optimization models that maximize matching quality subject to reviewer load constraints. However, this classical formulation is increasingly inadequate for modern conferences.

Large computer science venues now routinely receive more than $10{,}000$ submissions per cycle \cite{ZhaoZhang2022}, pushing paper assignment to unprecedented scales. At the same time, partially transparent review systems face growing risks of collusion and strategic manipulation: reviewers may exploit predictable assignment rules and manipulate bids to influence outcomes, which warrants robust assignment mechanisms \cite{PLRA,PM,AAAI2021}. Beyond robustness, program chairs often wish to promote diversity across institutions, regions, or career stages to ensure balanced evaluations. These requirements pose significant challenges for existing assignment methods.

Prior work has explored randomized and constrained approaches to address these issues. Jecmen et al.~\cite{PLRA} introduced PLRA, which mitigates collusion by computing a probabilistic assignment and sampling a deterministic matching with bounded pairwise probabilities. Xu et al.~\cite{PM} generalized this idea via a perturbed maximization objective that further spreads assignment probability across strong reviewers. While effective at introducing randomness, these methods incorporate diversity and anti-collusion only implicitly. In contrast, Leyton-Brown et al.~\cite{AAAI2021} proposed explicit linear constraints to prevent collusion and enforce diversity, but the resulting mixed-integer linear program is computationally prohibitive at conference scale.

In this paper, we propose a unified paper-assignment framework, Robust Assignment via Marginal Perturbation (RAMP). which is scalable, robust to collusion, and explicitly diversity-aware. Our approach combines a concave perturbed-maximization objective with soft constraints to compute a probabilistic assignment, a piecewise-linear approximation to enable efficient optimization, and a marginal-preserving, attribute-aware sampling procedure to produce a feasible deterministic assignment. The resulting pipeline achieves strong matching quality while scaling to large conference workloads.

Table \ref{tab:algorithm_properties} compares our algorithm with existing methods across several key properties, including runtime efficiency (Time), the use of randomness in the assignment process (Random), and practical considerations such as avoiding assignments to coauthors of the same paper (Coauthor), promoting geographical diversity (Div.), and eliminating 2-cycles (2Cyc.), which are situations in which two people bid for each other’s papers and are subsequently assigned to review them.

\begin{table}[H]
\centering
\small
\setlength{\tabcolsep}{3pt}        
\renewcommand{\arraystretch}{1.1}  

\begin{tabular}{@{}p{0.30\columnwidth}ccccc@{}}
\toprule
\textbf{Alg.} & \textbf{Time} & \textbf{Random} & \textbf{Coauthor} & \textbf{Div.} & \textbf{2Cyc.} \\
\midrule
Default & \checkmark & $\times$ & $\times$ & $\times$ & $\times$ \\
MILP~\cite{AAAI2021} & $\times$ & $\times$ & \checkmark & \checkmark & \checkmark \\
PLRA~\cite{PLRA} & \checkmark & \checkmark & $\times$ & $\times$ & $\times$ \\
PM~\cite{PM} & \checkmark$^{\ast}$ & \checkmark$^{\dagger}$ & $\times$ & $\times$ & $\times$ \\
RAMP & \checkmark & \checkmark$^{\dagger}$ & \checkmark & \checkmark & \checkmark \\
\bottomrule
\end{tabular}

\vspace{0.5ex}
{\footnotesize $^{\ast}$ partially;\;\; $^{\dagger}$ improved.}

\caption{Comparison of properties satisfied by different paper-assignment algorithms}
\label{tab:algorithm_properties}
\end{table}

\section{Preliminaries}

Consider an academic conference with a set $\mathcal P$ of papers and a set $\mathcal R$ of reviewers.
Each paper $p\in\mathcal P$ requires $\ell_p$ reviews, and each reviewer $r\in\mathcal R$ can review up to $u_r$ papers. Given a nonnegative similarity matrix 
$S\in\mathbb{R}_{\ge 0}^{|\mathcal P|\times|\mathcal R|}$, we seek an assignment 
$x_{p,r}\in\{0,1\}$ indicating whether reviewer $r$ is assigned to paper $p$.
The classical assignment algorithm uses a simple linear program that maximizes total similarity:
\begin{align*}
\max_{x \in [0,1]^{|\mathcal P|\times|\mathcal R|}}
\quad & \sum_{p\in \mathcal P}\sum_{r\in \mathcal R} S_{p,r}\, x_{p,r} \\
\text{s.t.}\quad
& \sum_{r\in \mathcal R} x_{p,r} = \ell_p \qquad && \forall p\in \mathcal P, \\
& \sum_{p\in \mathcal P} x_{p,r} \le u_r \qquad && \forall r\in \mathcal R, \\
& x_{p,r} = 0 \qquad && \forall (p,r)\in \mathcal F ,
\end{align*}
where $\mathcal F$ denotes the set of paper--reviewer pairs with conflicts of interest. 
The first two constraints enforce per-paper demand and per-reviewer capacity, respectively. 
Since the constraint matrix is totally unimodular, the LP admits an integral optimum.

The PLRA method~\cite{PLRA} introduces randomness by treating the problem as a continuous LP that maximizes expected matching quality while constraining each marginal probability $x_{p,r}\leq Q$. An integral solution is then sampled according to these marginals using the Birkhoff--von Neumann decomposition \cite{Birkhoff1946,Neumann+1953+5+12,Budish2009IMPLEMENTINGRA}, thereby limiting any malicious reviewer’s probability of being assigned to a specific paper.

Perturbed Maximization (PM) \cite{PM} extends this idea by optimizing a concave-perturbed objective
\[
\sum_{p,r} S_{p,r} \cdot f(x_{p,r}),
\]
where \(f\) is non-decreasing, concave with $f(0)=0$ and a decreasing derivative, rewarding the first units of probability more than later ones. This encourages probability mass to spread across several strong reviewers per paper. However, diversity and anti-collusion constraints are handled only \emph{indirectly} through randomness, leaving room for explicit modeling.

\section{A Unified and Scalable Paper Assignment Framework}\label{sec:framework}

At a high level, our main algorithmic contribution is a \emph{unified} and \emph{scalable} optimization framework for paper assignment that cleanly incorporates a rich family of \emph{soft} constraints. We build on the similarity-based randomized assignment formulations proposed in the PLRA~\cite{PLRA} and PM~\cite{PM} lines of work, which optimize a concave perturbation of similarity under standard paper and reviewer load constraints. In our framework, this baseline similarity objective is taken as a given, and we focus on enriching it with additional objective components that encode conference-specific desiderata.

Concretely, the framework is parameterized by a baseline similarity objective together with a collection of soft-constraint components. Each component $k$ is specified by an additional objective term $\mathcal{O}^k$ and a feasible region $\Xi^k$ defined by linear constraints over the assignment variables $x$ and a set of auxiliary variables $s^k$. Program chairs can therefore flexibly encode constraints related to diversity or anti-collusion rules by adding, removing, or reweighting these components, without changing the core optimization machinery.

Within this general framework, our main contribution on the modeling side is a collection of concrete soft-constraint components that promote diversity and more explicit handling of collusion. In particular, we introduce three such components of the objective function: $\mathcal{O}^{\mathsf{div}}$ for reviewer diversity, $\mathcal{O}^{\mathsf{coauthor}}$ for co-authorship penalties, and $\mathcal{O}^{\mathsf{2cycle}}$ for penalizing bid-based 2-cycle violations. These components are illustrative rather than exhaustive; conference organizers may plug in further components (e.g., institutional balance, track balance) by introducing corresponding objective–constraint pairs $(\mathcal{O}^k, \Xi^k)$ into the framework. The specific forms of these constraints are inspired by the constraints that were originally enforced in the MILP formulation of Leyton-Brown et al.~\cite{AAAI2021}.

To make the resulting program scalable at modern conference sizes, we further introduce three complementary algorithmic ingredients beyond modeling: (i) a piecewise-linear approximation that turns the nonlinear concave objective into a linear program that can be solved efficiently, (ii) an attribute-aware sampling procedure that converts the fractional solution into an integral assignment while mitigating sampling-induced violations of soft-constraints, and (iii) a sparsification strategy that restricts the optimization to a small set of high-quality candidate paper--reviewer pairs, reducing the problem size by orders of magnitude.

\subsection{A Unified Mathematical Program for Paper Assignment}

We formulate paper assignment as the following optimization problem. For each pair $(p,r)\in\mathcal P\times\mathcal R$, let $x_{p,r}\in[0,Q]$ denote the (fractional) probability of assigning reviewer $r$ to paper $p$, where $Q\in(0,1]$ upper-bounds the marginal assignment probability. Each paper $p$ requires $\ell_p$ reviews, and each reviewer $r$ has capacity $u_r$. Let $f:[0,Q]\to\mathbb{R}$ be a fixed concave, nondecreasing perturbation function (e.g., as in~\cite{PM}). The overall program is
\begin{align}
\max_{\mathbf{x},\mathbf{s}}\quad
&
\sum_{p\in\mathcal P}\sum_{r\in\mathcal R}
S_{p,r}\, f(x_{p,r})
+ \sum_{k=1}^K \mathcal{O}^k(\mathbf{x},s^k)
\label{eq:obj}
\\[-0.2em]
\text{s.t.}\quad
& \sum_{r\in\mathcal R} x_{p,r} = \ell_p,
\qquad \forall p\in\mathcal P,
\label{eq:ell}
\\
& \sum_{p\in\mathcal P} x_{p,r} \le u_r,
\qquad \forall r\in\mathcal R,
\\
& (x, s^1, \dots, s^K) \in \Xi^1 \cap \cdots \cap \Xi^K .
\label{eq:soft-constraints}
\end{align}

Here the first term is the similarity-based matching objective, each $\mathcal{O}^k$ denotes a soft-objective component with auxiliary variables $s^k$, and $\Xi^k$ specifies the corresponding linear constraints.

If all soft-objective terms and constraints are removed, the formulation reduces to standard similarity-based matching. In particular, setting $f(x)=x$ and $Q=1$ yields the deterministic assignment, $f(x)=x$ with $Q<1$ recovers the PLRA relaxation~\cite{PLRA}, and choosing a concave perturbation $f$ as in~\cite{PM} yields the PM objective. In this sense, our unified program subsumes several existing approaches as special cases, while additional components naturally extending it to incorporate explicit diversity and anti-collusion objectives.

\paragraph{Reviewer diversity.}
Often times, program chairs would want some diversity in the reviewers. For instance, a conference might seek geographic diversity to ensure that each paper receives perspectives from different parts of the world. This has the added benefit of reducing the likelihood of collusion, since reviewers and authors are less likely to know each other if they are geographically separated. Other conferences, especially those that are interdisciplinary, might want each paper to receive reviews from different areas of expertise, so that subject-area diversity is captured.  Finally, some conferences may want diversity in reviewer seniority, so that each paper gets at least one senior reviewer while also including junior reviewers.

$\mathcal{O}^{\mathsf{div}}$ and its associated constraints capture reviewer diversity. Program chairs may wish each paper to receive reviews from a diverse set of reviewers, according to some notion of \emph{region}.

Let $\mathcal{G}$ denote the set of regions under consideration for diversity (e.g., geographic regions, areas of expertise, or levels of seniority). For each reviewer $r \in \mathcal{R}$, let
\[
\mathrm{region}(r) \in \mathcal{G}
\]
denote the region to which $r$ belongs. We introduce a reward parameter $\lambda_{\mathsf{div}} \ge 0$ that controls the balance between diversity and similarity. For each paper $p \in \mathcal{P}$ and each region $g \in \mathcal{G}$, we introduce a slack variable $s^{\mathsf{div}}_{p,g}$ and impose the following constraints:
\begin{align}
0 \le s^{\mathsf{div}}_{p,g} &\le 1, \label{eq:div-range}\\
s^{\mathsf{div}}_{p,g} &\le \sum_{r \in \mathcal{R}:\,\mathrm{region}(r)=g} x_{p,r}.
\label{eq:div-upper}
\end{align}
Intuitively, $s^{\mathsf{div}}_{p,g}$ captures up to one unit of review probability that paper $p$ receives from region $g$.

Finally, we add the following term to the overall objective:
\begin{equation}
\mathcal{O}^{\mathsf{div}} \defeq \sum_{p \in \mathcal{P}} \sum_{g \in \mathcal{G}} \lambda_{\mathsf{div}} \, s^{\mathsf{div}}_{p,g}.
\label{eq:div-obj}
\end{equation}
These variables and constraints together define $\Xi^{\mathsf{div}}$.




\paragraph{Co-authorship distance.}
The co-authorship component $\mathcal{O}^{\mathsf{co}}$ and its associated constraints encode co-authorship structure. Reviewers assigned to the same paper should ideally not have co-authored previously, nor be very close in the co-authorship graph, in order to ensure independence of opinions and reduce the risk of collusion.

We introduce a penalty parameter $\lambda_{\mathsf{co}} \ge 0$. For each reviewer $r \in \mathcal{R}$, define the (closed) co-authorship neighborhood
\[
\mathcal{N}(r) \defeq \{r\} \cup \{ r' \in \mathcal{R} \mid r \text{ and } r' \text{ are past coauthors} \}.
\]

For each paper $p \in \mathcal{P}$ and reviewer $r \in \mathcal{R}$, we introduce a slack variable $s^{\mathsf{co}}_{p,r}$ and impose
\begin{align}
s^{\mathsf{co}}_{p,r} &\ge 1, \label{eq:coau-lb1}\\
s^{\mathsf{co}}_{p,r} &\ge \sum_{r' \in \mathcal{N}(r)} x_{p,r'} .
\label{eq:coau-lb2}
\end{align}
The second constraint ensures that $s^{\mathsf{co}}_{p,r}$ is lower bounded by the total assignment probability allocated to reviewer $r$ and all reviewers in $r$’s co-authorship neighborhood.

We then add the following term to the objective:
\begin{equation}
\mathcal{O}^{\mathsf{co}} \defeq - \sum_{p \in \mathcal{P}} \sum_{r \in \mathcal{R}} \lambda_{\mathsf{co}} \, s^{\mathsf{co}}_{p,r}.
\end{equation}
These variables and constraints together define $\Xi^{\mathsf{co}}$.

This penalty discourages assigning direct past coauthors to the same paper and also indirectly penalizes reviewer pairs at co-authorship distance~2, since such pairs share a common reviewer in their neighborhoods. As a result, it helps avoid assigning reviewers drawn from tightly connected subcommunities.

Moreover, we observe empirically that applying this penalty only to reviewer--paper pairs $(p,r)$ for which reviewer $r$ submits a positive bid on paper $p$ substantially sparsifies the optimization problem without noticeably degrading solution quality, and we adopt this strategy in our implementation.

\paragraph{2-cycles.}
The 2-cycle component $\mathcal{O}^{\mathsf{cyc}}$ penalizes bid-based 2-cycle violations. Ideally, pairs of reviewers who bid positively on each other’s papers should not be assigned to review those papers, providing an additional safeguard against bid-based collusion.

Let $\mathcal{C}$ denote the set of all 2-cycles $(r_1,r_2,p_1,p_2)$ such that reviewer $r_1$ bids positively on reviewer $r_2$’s paper $p_2$ and reviewer $r_2$ bids positively on reviewer $r_1$’s paper $p_1$. We introduce a penalty parameter $\lambda_{\mathsf{cyc}} \ge 0$ and add the following term to the objective:
\begin{equation}
\mathcal{O}^{\mathsf{cyc}} \defeq
- \sum_{(r_1,r_2,p_1,p_2)\in \mathcal{C}}
\lambda_{\mathsf{cyc}} \, \bigl(x_{p_2,r_1} + x_{p_1,r_2}\bigr)^2.
\label{eq:2cycle}
\end{equation}
The auxiliary variables introduced when linearizing this quadratic term, together with their associated linear constraints, are included in $\Xi^{\mathsf{cyc}}$.

The quadratic expression $\bigl(x_{p_2,r_1} + x_{p_1,r_2}\bigr)^2$ is convex, and therefore the penalization term
$-\lambda_{\mathsf{cyc}} \bigl(x_{p_2,r_1} + x_{p_1,r_2}\bigr)^2$
is concave, preserving the overall concavity of the objective.

Expanding the square yields
\[
\bigl(x_{p_2,r_1} + x_{p_1,r_2}\bigr)^2
=
x_{p_2,r_1}^2
+ 2 x_{p_2,r_1} x_{p_1,r_2}
+ x_{p_1,r_2}^2,
\]
where
\begin{itemize}
    \item the cross term $2 x_{p_2,r_1} x_{p_1,r_2}$ captures the probability that both assignments occur simultaneously, which is precisely the behavior we seek to penalize;
    \item the remaining squared terms depend only on individual assignment variables and introduce no additional interactions. These terms can be absorbed into the per-assignment perturbed-maximization objective $f$, increasing the penalty on large marginal assignment probabilities for reviewer--paper pairs involved in a 2-cycle without altering the structure of the optimization problem.
\end{itemize}

\paragraph{Seniority coverage.}
At the scale of large modern-day conferences, program chairs may require that each paper receive at least one experienced reviewer to ensure the quality and reliability of the reviews. While this objective could in principle be implemented in a manner similar to the geographic diversity constraints above, it is often treated as a hard requirement that should be satisfied with a $100\%$ guarantee. Accordingly, our framework is able to support this case via a simple two-stage procedure: we run the matching algorithm twice, first on the set of senior reviewers only to guarantee coverage, and then again on the set of junior reviewers to fill the remaining slots. We report experimental results for this strategy in \Cref{sec:appendix-sen}.

\subsection{Piecewise Linearization of the Optimization Program}
\label{subsec:ablation_piecewise_linear_and_sampling}

As in PM~\cite{PM}, the objective contains nonlinear concave terms, including the perturbed similarity terms $S_{p,r} f(x_{p,r})$ and the concave quadratic penalties in $\mathcal{O}^{\mathsf{cyc}}$, which reduce computational efficiency. To improve scalability, we approximate each concave function by a piecewise-linear surrogate.

Specifically, for a concave function $f$ defined on $[0,Q]$, we construct a piecewise-linear approximation $g$ using uniform segments of length $\Delta>0$ via linear interpolation between grid points $\{k\Delta\}$, where $k=0,1,2,\ldots,\lfloor Q/\Delta \rfloor$. As $\Delta \to 0$, continuity of $f$ implies that $g$ converges uniformly to $f$ on $[0,Q]$.

The resulting objective is separable and concave piecewise-linear. Since any one-dimensional concave piecewise-linear function can be expressed as the minimum of finitely many affine functions, there exist coefficients $\{a_j,b_j\}_{j\in J}$ such that
\[
g(x)=\min_{j\in J}\{a_j x+b_j\}.
\]
Introducing auxiliary variables $t_{p,r}$ with linear constraints
\[
t_{p,r}\le a_j x_{p,r}+b_j,\quad \forall j\in J,
\]
allows us to replace each nonlinear term by an equivalent linear formulation. An analogous construction is applied to the concave quadratic terms in $\mathcal{O}^{\mathsf{cyc}}$. All constraints remain linear, and the resulting approximation is therefore a linear program.

In our implementation, we set $\Delta=0.1$. An ablation study of this approximation is provided in~\Cref{subsec:ablation_piecewise_linear_and_sampling}.

\subsection{Attribute-Aware Sampling}\label{subsec:sampling}

In the previous subsection, the optimization program produces a fractional (probabilistic) assignment $x_{p,r}$. The next step is to generate a final integral assignment based on these marginal probabilities. Following~\cite{PLRA,PM}, we employ the Birkhoff--von Neumann (BVN) decomposition for sampling, applied to the bipartite graph induced by the nonzero entries of $x$.

Among our soft constraints, those induced by $\mathcal{O}^{\mathsf{div}}$ and $\mathcal{O}^{\mathsf{coauthor}}$ are specifically designed to promote diversity and reduce collusion. However, the stochastic nature of sampling can lead to additional violations of these constraints. For example, if two reviewers from the same region each have probability $0.5$ of being assigned to a paper $p$, both may end up being selected after sampling.

To mitigate such effects, we modify the BVN sampling procedure. The BVN algorithm iteratively finds a loop
\[
(r_0\rightarrow p_0\rightarrow r_1\rightarrow p_1\rightarrow\cdots\rightarrow r_{\ell-1}\rightarrow p_{\ell-1}\rightarrow r_0),
\]
where, for each $i\in\{0,1,\dots,\ell-1\}$, both $x_{p_i,r_i}$ and $x_{p_i,r_{(i+1)\bmod \ell}}$ remain fractional. A step size $\Delta$ is then drawn in the same manner as in Algorithm~1 of~\cite{PLRA}. The value of $\Delta$ may be positive or negative. Each $x_{p_i,r_i}$ is updated by $+\Delta$ and each $x_{p_i,r_{(i+1)\bmod \ell}}$ is updated by $-\Delta$. After these adjustments are applied to all edges in the loop, the algorithm proceeds to the next iteration.

The original BVN algorithm does not specify how to select the loop at each iteration; any valid loop suffices for correctness. The simplest implementation, as in~\cite{PLRA}, uses a depth-first search (DFS) to find such a loop. In our variant, we bias this DFS toward improving diversity. When extending a partial loop 
\[
(r_0\rightarrow p_0\rightarrow r_1\rightarrow p_1\rightarrow\dots\rightarrow r_{k}\rightarrow p_{k}),
\]
and searching for the next reviewer $r_{k+1}$ adjacent to $p_k$, we proceed as follows:
\begin{enumerate}
    \item First, we give priority to reviewers $r_{k+1}$ who are past coauthors of $r_k$ (i.e., $r_{k+1}\in \mathcal{N}(r_k)$);
    \item If no such reviewer is available, we prioritize reviewers from the same diversity region as $r_k$ (i.e., $\mathrm{region}(r_{k+1}) = \mathrm{region}(r_{r_k})$).
\end{enumerate}
We prioritize co-authorship distance over region similarity because, intuitively, two past coauthors are more similar than two reviewers who merely share a region (e.g., geographic location).


This biased sampling method naturally reduces the likelihood that two similar reviewers are assigned to the same paper. To illustrate this, consider again the example of two reviewers from the same region, each having a probability of $0.5$ of being assigned to paper~$p$. One approach is to increase one reviewer’s assignment probability by $0.5$ while decreasing the other’s by $0.5$. This ensures that exactly one of the two reviewers is assigned to the paper, and therefore no diversity violation can occur. By contrast, suppose we instead increase or decrease both reviewers’ probabilities by $0.5$ together, each with probability $0.5$. A diversity violation can occur only in the case where both probabilities are increased, so we have that the constraint is violated with probability $1/4$


The pseudocode of the sampling algorithm is presented in \Cref{app:sampling_algo}. An ablation study evaluating this sampling strategy is presented in \Cref{subsec:ablation_piecewise_linear_and_sampling}. Note that if conference organizers wish to impose additional diversity constraints, these can be incorporated into the same sampling framework with minimal modification.

\subsection{Scalable Implementation via Sparsification} \label{subsec:sparse}

In large conferences, the similarity matrix $S$ is extremely sparse in terms of ``useful'' entries. To exploit this, our algorithm restricts attention to a sparse subset of candidate paper--reviewer pairs.

Concretely, for each paper $p\in\mathcal{P}$, we retain only the top $K_{\text{paper}}$ reviewers ranked by $S_{p,r}$, and for each reviewer $r\in\mathcal{R}$, we retain only the top $K_{\text{rev}}$ papers. The union of these pairs defines the support of the assignment variables $x_{p,r}$; all remaining pairs are removed from the optimization program.

In our experiments, we set $K_{\text{paper}} = K_{\text{rev}} = 1000$. For a conference-scale instance with roughly $20{,}000$ papers and reviewers, this yields a highly sparse problem in which the number of active variables is several orders of magnitude smaller than $|\mathcal{P}|\cdot|\mathcal{R}|$. Accordingly, we include only these variables and the constraints involving them in the solver.

By comparison, the NeurIPS 2024/2025 paper-matching system applies a similar sparsification to similarity scores but retains all variables in the optimization, treating non-retained pairs as zero-weighted entries. Explicitly removing these variables leads to substantial reductions in both memory usage and runtime in our setting.

\section{Experiments}\label{sec:exp}

\subsection{Experimental Setup}

\paragraph{Algorithms.} We evaluate the following algorithms: \textbf{Default}, The default LP algorithm; \textbf{MILP:} The mixed-integer linear program algorithm as described in \cite{AAAI2021}; \textbf{PLRA}, The Probability-Limited Randomized Assignment algorithm in \cite{PLRA}; \textbf{PM}, The Perturbed-Maximization algorithm in \cite{PM}. 
Lastly, we have \textbf{RAMP}, our proposed algorithm.

For the MILP algorithm, we adopt the hyperparameters and row-generation settings from the AAAI 2021 implementation \cite{AAAI2021}. For PLRA, PM, and RAMP, we set the parameter $Q = 0.9$. Additionally, for both PM and our algorithm, we use the quadratic version of the concave function $f(x) = x - 0.1x^2$, as used in NeurIPS 2024 and 2025.

\paragraph{Datasets.} We evaluate our algorithm on four datasets summarized in Table~\ref{tab:datasets}. The Large dataset is a synthetic conference-scale instance; full generation details are provided in the \Cref{app:details-synthetic-data}. We additionally use two smaller real-world datasets from AAMAS~2015 and ICLR~2018, previously studied in~\cite{PM}, as well as the S2ORC dataset \cite{S2ORC}, which models
bid-based collusion using citation-derived signals.
For datasets lacking reviewer metadata or bids, we generate these attributes using the same procedure as for the Large dataset.

\begin{table}[h]
\centering
\small
\setlength{\tabcolsep}{6pt}
\renewcommand{\arraystretch}{1.05}
\begin{tabular}{lccc}
\toprule
\textbf{Dataset} & \textbf{\# Papers} & \textbf{\# Reviewers} & \textbf{Source} \\
\midrule
Large & 20{,}000 & 22{,}000 & Synthetic \\
AAMAS 2015 & 601 & 213 & Real \\
ICLR 2018 & 911 & 2{,}435 & Real \\
S2ORC & 2{,}446 & 2{,}483 & Synthetic\\
\bottomrule
\end{tabular}
\caption{Datasets used in evaluation}
\label{tab:datasets}
\end{table}





\paragraph{Metrics.} To assess the performance of our algorithm, we use the following evaluation metrics:

\begin{itemize}
    \item \textbf{Quality:} The objective of the algorithm relative to the best possible objective value (which is usually achieved by the default algorithm).
\item \textbf{Runtime:} The time required to execute the algorithm, reflecting its computational efficiency.
\item \textbf{Diversity:} Calculated as the average number of distinct regions assigned to each paper by all reviewers.
\item \textbf{Coauthors:} the total number of coauthor pairs assigned to the same paper.
\item \textbf{2-Cycles:} the number of 2-cycles as defined in previous sections.
\end{itemize}

Similar to \cite{PM}, we use two randomness-related metrics to evaluate the variability and diversity of the assignments: 

\begin{itemize}
    \item \textbf{Support Size:}
$\mathrm{Support}(\mathbf{x}) = \sum_{p,r} \mathbb{I}[x_{p,r} > 0]$,
which counts the number of positive assignments. Maximizing support size enhances the estimation quality across multiple valid paper--reviewer matchings.
\item \textbf{Entropy:}
$\mathrm{Entropy}(\mathbf{x}) = -\sum_{p,r} x_{p,r} \ln(x_{p,r})$,
which encourages uncertainty and diversity in the assignments, promoting a more balanced and randomized allocation.
\end{itemize} 

\paragraph{Implementation Details.} 


All experiments were conducted on an Amazon Web Services (AWS) EC2 r6i.4xlarge instance equipped with 16 virtual CPUs (vCPUs) and 128 GiB of memory. For all algorithms, we use the commercial solver Gurobi \cite{gurobi}, following \cite{PM}, to solve the corresponding mathematical optimization programs.

The sparsification technique described in Section \ref{subsec:sparse} was applied to all evaluated algorithms, as otherwise even machines with up to 512 GiB of memory would encounter memory exhaustion when running Gurobi, making large-scale experiments infeasible. Furthermore, note that in our experiments negative similarity scores are treated as Conflicts Of Interests.

\subsection{Comparison with Prior Methods}


\Cref{tab:large} compares RAMP against four baselines on the Large dataset. RAMP is the only method that achieves substantial improvements over Default across all robustness-related metrics as well as all diversity-related metrics, including coauthor pairs, two-cycles, and geographic diversity. 

In addition, RAMP completes the large-scale experiment in under 20 minutes, representing an 89\% reduction in runtime compared to MILP, which is the only baseline that also performs strongly across all robustness-related metrics. These gains are achieved while incurring only a 2.6\% reduction in assignment Quality compared to Default. These trends are consistent across datasets, with results on additional datasets reported in \cref{app:other_datasets}.

\begin{table}[t]
\centering
\small
\setlength{\tabcolsep}{4pt}
\renewcommand{\arraystretch}{1.1}

\begin{tabular}{@{}lcccc|c@{}}
\toprule
\textbf{Metric                                                                                                                               } 
& \textbf{Default} 
& \textbf{MILP} 
& \textbf{PM} 
& \textbf{PLRA} 
& \textbf{RAMP} \\
\midrule
Time (s) $\downarrow$     & 585.0   & 17257.5 & 1765.4 & 600.1  & 1096.1 \\
Quality $\uparrow$      & 1.000   & 0.985   & 0.980 & 0.999 & 0.974 \\
Support $\uparrow$      & 80000   & 80000   & 779680 & 103576 & 537266 \\
Entropy $\uparrow$      & 0       & 0       & 155316 & 16355  & 143587 \\
Div.   $\uparrow$       & 0.615   & 0.916   & 0.616  & 0.614  & 0.895 \\
Coauth. $\downarrow$      & 163     & 0       & 150    & 172    & 21 \\
2Cyc.  $\downarrow$      & 86      & 0       & 117    & 86     & 1 \\
\bottomrule
\end{tabular}

\caption{Performance comparison of paper-assignment algorithms on the large synthetic dataset. Metrics marked with $\uparrow$ indicate that higher values are preferred, while those marked with $\downarrow$ indicate that lower values are preferred.}
\label{tab:large}
\end{table}


\subsection{Hyperparameters, Quality, and Constraint Satisfaction}

One natural question is: how do the choices of hyperparameters for our soft constraints impact key attributes such as quality, runtime, and how well different constraints are satisfied? 
\Cref{tab:hyperparam-comparison} illustrates the trade-offs induced by tuning hyperparameters $\lambda_*$ in soft constraints on S2ORC: we start with all hyperparameters set to $0$, and tune individual hyperparameters.  Increasing individual penalties improves the targeted constraint with only a minor impact on quality and runtime. For example, increasing $\pencycle$ eliminates two-cycles with less than a 0.2\% quality loss. Additional sweeps are provided in \cref{sec:hyperparam}.





\begin{table}[h]
\centering
\small
\setlength{\tabcolsep}{3pt}
\renewcommand{\arraystretch}{1.1}

\begin{tabular}{@{}lccccc@{}}
\toprule
Setting & Time (s) $\downarrow$ & Quality $\uparrow$ & Div. $\uparrow$ & Coauth. $\downarrow$ & 2Cyc. $\downarrow$\\
\midrule
$\text{all} = 0.0$                & 113.0 & 1.000 & 0.835 & 76 & 101 \\
$\rewarddiv = 0.15$      & 125.0 & 0.994 & \textbf{0.949} & 55 & 88 \\
$\pencoauthor = 0.15$    & 113.0 & 0.999 & 0.832 & \textbf{30} & 104 \\
$\pencycle = 0.2$        & 112.3 & 0.998 & 0.846 & 80 & \textbf{0} \\
\bottomrule
\end{tabular}

\caption{Comparison on the S2ORC dataset, with all parameters set to $0.0$ except for one tuned to its preferred value}
\label{tab:hyperparam-comparison}
\end{table}

\subsection{Ablation Studies}\label{subsec:ablation_piecewise_linear_and_sampling}

\paragraph{Piecewise-linear Approximation.} To assess the contribution of the piecewise-linear (PWL) approximation, we compare our full algorithm against a variant that operates directly on the nonlinear objective and constraint functions, without any linearization. Both versions were evaluated on the Large dataset, using identical hyperparameter settings to ensure a fair comparison. The quantitative results are presented in \Cref{tab:pwl_ablation}, with only the most relevant attributes. The full table with all attributes is displayed in the \cref{app:ablation}. 


\begin{table}[h]
\centering
\small
\setlength{\tabcolsep}{4pt}
\renewcommand{\arraystretch}{1.1}

\begin{tabular}{@{}lccc@{}}
\toprule
\textbf{Variant} & \textbf{Time (s)} $\downarrow$ & \textbf{Quality} $\uparrow$ & \textbf{Support} $\uparrow$ \\
\midrule
PWL      & 1088.9 & 1.0 & 537266 \\
Non-PWL  & 16569.9 & 0.999 & 825421 \\
\bottomrule
\end{tabular}

\caption{Comparison of the PWL approximation and the non-PWL variant on the large synthetic dataset}
\label{tab:pwl_ablation}
\end{table}

\Cref{tab:pwl_ablation} shows that removing the PWL approximation increases runtime by over $16\times$, while leaving assignment quality and constraint metrics essentially unchanged. This confirms that PWL is critical for computational efficiency without sacrificing solution quality.


\paragraph{Attribute-aware sampling algorithm.} We examine the impact of the attribute-aware sampling algorithm introduced in \Cref{subsec:sampling}. We conducted experiments on the Large dataset, comparing our method with and without attribute-aware sampling enhancement. Similarly, both variants were run under identical hyperparameter settings to ensure a controlled comparison. \Cref{tab:sampling_ablation} summarizes the results across several key metrics, while other metrics showed no significant difference; the complete table is presented in \Cref{app:ablation}. 


\begin{table}[h]
\centering
\small
\setlength{\tabcolsep}{2pt}
\renewcommand{\arraystretch}{1.1}

\begin{tabular}{@{}lccccc@{}}
\toprule
\textbf{Method} & \textbf{Time (s)} $\downarrow$ & \textbf{Quality} $\uparrow$ & \textbf{Div.} $\uparrow$ & \textbf{Coauth.} $\downarrow$ \\
\midrule
Attribute-Aware       & 1088.9 & 1.0 & \textbf{0.896} & \textbf{26} \\
Vanilla & 1083.2 & 1.0 & 0.697 & 199 \\
\bottomrule
\end{tabular}
\caption{Comparison of different sampling methods on the large synthetic dataset}
\label{tab:sampling_ablation}
\end{table}

As shown in \Cref{tab:sampling_ablation}, attribute-aware sampling leaves runtime and quality unchanged, but substantially improves some diversity-related metrics. In particular, geographic diversity increases from $0.697$ to $0.896$, and coauthor pairs drop from $199$ to $26$, demonstrating that the sampling procedure effectively enforces diversity constraints without affecting efficiency.

\section{Related Work}\label{sec:rel_works}

This paper builds on a growing literature on algorithmic paper assignment for peer review. Starting from maximum-similarity matching, prior work has introduced extensions addressing distributional guarantees, incentives, and robustness \cite{pmlr-v98-stelmakh19a,JecmenZhangLiuFangConitzerShah2022,DhullJecmenKothariShah2022,S2ORC}. Our work is most closely related to optimization-based formulations and randomized assignment methods for robustness.

Reviewer assignment is typically formulated as an optimization problem maximizing total matching quality subject to capacity and conflict-of-interest constraints \cite{flach2010sigkdd,garg2010assigning,tang2010expertise,lian2018conference}, which reduces to a polynomial-time network-flow problem in the absence of constraints \cite{taylor2008optimal}. To address inequities, Stelmakh et al.~\cite{pmlr-v98-stelmakh19a} propose max–min fairness via flow-based relaxations where they focused on finding matches that maximize the minimum matching, while Kobren et al.~\cite{kobren2019papermatching} incorporate local balance and load-bounding constraints within an integer linear program.

Randomized assignment has been proposed to reduce predictability and defend against strategic behavior and collusion \cite{jecmen2022tradeoffs,jecmen2023dataset}. Methods such as PLRA and PM introduce controlled randomness \cite{PLRA,PM}, while other approaches model bidding behavior or eliminate short cycles to prevent quid-pro-quo arrangements \cite{S2ORC,boehmer2022collusion,AAAI2021}. Randomization has also been used to study bias, herding, and to enable counterfactual evaluation of assignment policies \cite{tomkins2017bias,lawrence2014nips,stelmakh2023herding,saveski2023counterfactual}.

Our use of piecewise linearization is closely related to the approach of Xu et al.~\cite{PM}, who propose a similar approximation to the perturbed-maximization objective. In their setting, the authors interpret the piecewise-linear approximation through a network-flow formulation and empirically observe that directly solving the original quadratic program with Gurobi can be competitive. In contrast, we adopt the same piecewise-linearization idea but directly solve the resulting linear program, which proves to be more efficient in our setting.

\section{Acknowledgment}

This work is supported by NSF IIS-2200410. We also thank Matthew Taylor, Chad Jenkins, and Kevin Leyton-Brown for their valuable input and feedback.




\bibliographystyle{alpha}
\bibliography{ref}

@misc{gurobi,
  author = {{Gurobi Optimization, LLC}},
  title  = {Gurobi Optimizer},
  label  = {Gurobi},
  key    = {Gurobi},
  year   = {2024},
  note   = {\url{https://www.gurobi.com}}
}

@article{S2ORC,
  title={Making Paper Reviewing Robust to Bid Manipulation Attacks},
  author={Ruihan Wu and Chuan Guo and Felix Wu and Rahul Kidambi and Laurens van der Maaten and Kilian Q. Weinberger},
  journal={ArXiv},
  year={2021},
  volume={abs/2102.06020},
  url={https://api.semanticscholar.org/CorpusID:231879710}
}

@inproceedings{Budish2009IMPLEMENTINGRA,
  title={IMPLEMENTING RANDOM ASSIGNMENTS : A GENERALIZATION OF THE BIRKHOFF-VON NEUMANN THEOREM},
  author={Eric Budish and Yeon-Koo Che and Fuhito Kojima and Paul R. Milgrom},
  year={2009},
  url={https://api.semanticscholar.org/CorpusID:11068959}
}

@inbook{Neumann+1953+5+12,
url = {https://doi.org/10.1515/9781400881970-002},
title = {1. A Certain Zero-sum Two-person Game Equivalent to the Optimal Assignment Problem},
booktitle = {Contributions to the Theory of Games, Volume II},
author = {John von Neumann},
editor = {Harold William Kuhn and Albert William Tucker},
publisher = {Princeton University Press},
address = {Princeton},
pages = {5--12},
doi = {doi:10.1515/9781400881970-002},
isbn = {9781400881970},
year = {1953},
lastchecked = {2025-10-09}
}

@article{Birkhoff1946,
  author    = {Garrett Birkhoff},
  title     = {Three Observations on Linear Algebra},
  journal   = {Univ. Nac. Tucum{\'a}n. Rev. Ser. A},
  volume    = {5},
  pages     = {147--151},
  year      = {1946}
}

@inproceedings{jecmen2022tradeoffs,
  title={Tradeoffs in preventing manipulation in paper bidding for reviewer assignment},
  author={Jecmen, Steven and Shah, Nihar B and Fang, Fei and Conitzer, Vincent},
  booktitle={ML Evaluation Standards Workshop at ICLR},
  year={2022}
}

@inproceedings{jecmen2023dataset,
  title={A dataset on malicious paper bidding in peer review},
  author={Jecmen, Steven and Yoon, Minji and Conitzer, Vincent and Shah, Nihar B. and Fang, Fei},
  booktitle={Proceedings of the Web Conference (WWW)},
  year={2023}
}

@inproceedings{boehmer2022collusion,
  title={Combating collusion rings is hard but possible},
  author={Boehmer, Niclas and Bredereck, Robert and Nichterlein, Andr{\'e}},
  booktitle={Proceedings of the AAAI Conference on Artificial Intelligence},
  year={2022}
}

@article{tomkins2017bias,
  title={Reviewer bias in single- versus double-blind peer review},
  author={Tomkins, Andrew and Zhang, Min and Heavlin, William D.},
  journal={Proceedings of the National Academy of Sciences},
  volume={114},
  number={48},
  pages={12708--12713},
  year={2017}
}

@inproceedings{saveski2023counterfactual,
  title={Counterfactual Evaluation of Peer Review Assignment Strategies in Computer Science and Artificial Intelligence},
  author={Saveski, Martin and Jecmen, Steven and Shah, Nihar and Ugander, Johan},
  booktitle={Advances in Neural Information Processing Systems (NeurIPS)},
  year={2023}
}

@misc{lawrence2014nips,
  title={The NIPS experiment},
  author={Lawrence, Neil D.},
  howpublished={\url{https://inverseprobability.com/2014/12/16/the-nips-experiment}},
  note={Accessed February 1, 2023},
  year={2014}
}

@article{flach2010sigkdd,
  title={Novel tools to streamline the conference review process: Experiences from SIGKDD'09},
  author={Flach, Peter A. and Spiegler, Sebastian and Gol{\'e}nia, Bruno and Price, Simon and Guiver, John and Herbrich, Ralf and Graepel, Thore and Zaki, Mohammed J.},
  journal={ACM SIGKDD Explorations Newsletter},
  volume={11},
  number={2},
  pages={63--67},
  year={2010}
}

@article{garg2010assigning,
  title={Assigning papers to referees},
  author={Garg, Naveen and Kavitha, Telikepalli and Kumar, Amit and Mehlhorn, Kurt and Mestre, Juli{\'a}n},
  journal={Algorithmica},
  volume={58},
  number={1},
  pages={119--136},
  year={2010},
  doi={10.1007/s00453-009-9386-0},
  url={https://doi.org/10.1007/s00453-009-9386-0}
}

@inproceedings{tang2010expertise,
  title={Expertise matching via constraint-based optimization},
  author={Tang, Wenbin and Tang, Jie and Tan, Chenhao},
  booktitle={Proceedings of the IEEE/WIC/ACM International Conference on Web Intelligence},
  pages={34--41},
  year={2010},
  doi={10.1109/WI-IAT.2010.133},
  url={https://doi.org/10.1109/WI-IAT.2010.133}
}

@techreport{taylor2008optimal,
  title={On the Optimal Assignment of Conference Papers to Reviewers},
  author={Taylor, Camillo J.},
  year={2008}
}

@inproceedings{lian2018conference,
  title={The Conference Paper Assignment Problem: Using Order Weighted Averages to Assign Indivisible Goods},
  author={Lian, Jing Wu and Mattei, Nicholas and Noble, Renee and Walsh, Toby},
  booktitle={Proceedings of the AAAI Conference on Artificial Intelligence},
  pages={1138--1145},
  year={2018},
  url={https://www.aaai.org/ocs/index.php/AAAI/AAAI18/paper/view/17396}
}

@inproceedings{kobren2019papermatching,
  title={Paper Matching with Local Fairness Constraints},
  author={Kobren, Ari and Saha, Barna and McCallum, Andrew},
  booktitle={Proceedings of the ACM SIGKDD International Conference on Knowledge Discovery \& Data Mining (KDD)},
  pages={1247--1257},
  year={2019},
  doi={10.1145/3292500.3330899},
  url={https://doi.org/10.1145/3292500.3330899}
}

@article{stelmakh2023herding,
  title={A large scale randomized controlled trial on herding in peer-review discussions},
  author={Stelmakh, Ivan and Rastogi, Charvi and Shah, Nihar B and Singh, Aarti and Daum{\'e}, Hal III},
  journal={PLoS ONE},
  year={2023}
}

@inproceedings{PM,
author = {Xu, Yixuan Even and Jecmen, Steven and Song, Zimeng and Fang, Fei},
title = {A one-size-fits-all approach to improving randomness in paper assignment},
year = {2023},
publisher = {Curran Associates Inc.},
address = {Red Hook, NY, USA},
booktitle = {Proceedings of the 37th International Conference on Neural Information Processing Systems},
articleno = {636},
numpages = {24},
location = {New Orleans, LA, USA},
series = {NIPS '23}
}

@article{AAAI2021,
author = {Leyton-Brown, Kevin and Mausam and Nandwani, Yatin and Zarkoob, Hedayat and Cameron, Chris and Newman, Neil and Raghu, Dinesh},
title = {Matching papers and reviewers at large conferences},
year = {2024},
issue_date = {Jun 2024},
publisher = {Elsevier Science Publishers Ltd.},
address = {GBR},
volume = {331},
number = {C},
issn = {0004-3702},
url = {https://doi.org/10.1016/j.artint.2024.104119},
doi = {10.1016/j.artint.2024.104119},
journal = {Artif. Intell.},
month = jun,
numpages = {24}
}

@inproceedings{PLRA,
author = {Jecmen, Steven and Zhang, Hanrui and Liu, Ryan and Shah, Nihar B. and Conitzer, Vincent and Fang, Fei},
title = {Mitigating manipulation in peer review via randomized reviewer assignments},
year = {2020},
isbn = {9781713829546},
publisher = {Curran Associates Inc.},
address = {Red Hook, NY, USA},
booktitle = {Proceedings of the 34th International Conference on Neural Information Processing Systems},
articleno = {1051},
numpages = {13},
location = {Vancouver, BC, Canada},
series = {NIPS '20}
}

@CONFERENCE{Protasiewicz20143062,
	author = {Protasiewicz, Jarosław},
	title = {A support system for selection of reviewers},
	year = {2014},
	journal = {Conference Proceedings - IEEE International Conference on Systems, Man and Cybernetics},
	volume = {2014-January},
	number = {January},
	pages = {3062 – 3065},
	doi = {10.1109/smc.2014.6974397},
	url = {https://www.scopus.com/inward/record.uri?eid=2-s2.0-84938149994&doi=10.1109%2fsmc.2014.6974397&partnerID=40&md5=3ff45e78db8eedf0542261e958f63708},
	type = {Conference paper},
	publication_stage = {Final},
	source = {Scopus},
	note = {Cited by: 22; All Open Access, Gold Open Access}
}

@article{ZhaoZhang2022,
  title     = {Reviewer Assignment Algorithms for Peer Review Automation: A Survey},
  author    = {Zhao, Xiquan and Zhang, Yangsen},
  journal   = {Information Processing and Management},
  volume    = {59},
  number    = {6},
  pages     = {103028},
  year      = {2022},
  issn      = {0306-4573},
  doi       = {10.1016/j.ipm.2022.103028},
  url       = {https://doi.org/10.1016/j.ipm.2022.103028},
  publisher = {Elsevier},
  note      = {Open access under CC BY 4.0 license}
}

@InProceedings{pmlr-v98-stelmakh19a,
  title = 	 {PeerReview4All: Fair and Accurate Reviewer Assignment in  Peer Review},
  author =       {Stelmakh, Ivan and Shah, Nihar B. and Singh, Aarti},
  booktitle = 	 {Proceedings of the 30th International Conference on Algorithmic Learning Theory},
  pages = 	 {828--856},
  year = 	 {2019},
  editor = 	 {Garivier, Aurélien and Kale, Satyen},
  volume = 	 {98},
  series = 	 {Proceedings of Machine Learning Research},
  month = 	 {22--24 Mar},
  publisher =    {PMLR},
  url = 	 {https://proceedings.mlr.press/v98/stelmakh19a.html}
}

@InProceedings{JecmenZhangLiuFangConitzerShah2022,
  author    = {Jecmen, Steven and Zhang, Hanrui and Liu, Ryan and Fang, Fei and Conitzer, Vincent and Shah, Nihar B.},
  title     = {Near-Optimal Reviewer Splitting in Two-Phase Paper Reviewing and Conference Experiment Design},
  booktitle = {Proceedings of the 10th AAAI Conference on Human Computation and Crowdsourcing (HCOMP 2022)},
  year      = {2022},
  month     = {October},
  doi       = {10.1609/hcomp.v10i1.21991},
  url       = {https://ojs.aaai.org/index.php/HCOMP/article/view/21991},
  pages     = {1-13}
}

@InProceedings{DhullJecmenKothariShah2022,
  author    = {Dhull, Komal and Jecmen, Steven and Kothari, Pravesh and Shah, Nihar B.},
  title     = {Strategyproofing Peer Assessment via Partitioning: The Price in Terms of Evaluators’ Expertise},
  booktitle = {Proceedings of the 10th AAAI Conference on Human Computation and Crowdsourcing (HCOMP 2022)},
  pages     = {53--63},
  year      = {2022},
  editor    = {Hsu, Jane and Yin, Ming},
  volume    = {10},
  publisher = {Association for the Advancement of Artificial Intelligence},
  doi       = {10.1609/hcomp.v10i1.21987},
  url       = {https://ojs.aaai.org/index.php/HCOMP/article/view/21987}
}

\newpage
\clearpage
\appendix

\section{Details on Synthetic Data Generation}
\label{app:details-synthetic-data}

We generated a large synthetic dataset as follows. This dataset, along with all other datasets used in our experiments, will be released on GitHub upon publication. Each paper and reviewer was assigned to one of five topical areas according to a skewed distribution, designed to reflect the topical imbalance observed in real-world conferences.

To model subject-matter affinity, we computed a dense similarity score for every paper--reviewer pair using latent paper- and reviewer-specific parameters. These scores were adjusted for topical alignment and further modulated by simulated bid signals. The resulting affinity scores had a mean of approximately $0.6$ and were clipped to the interval $[0,1]$. To obtain a computationally tractable representation, we retained only the top $1{,}000$ reviewer matches per paper and the top $1{,}000$ paper matches per reviewer, yielding a truncated but high-quality similarity matrix for downstream experiments. This sparsification procedure follows the method described in \Cref{subsec:sparse}. After truncation, the mean affinity score increased to approximately $0.8$.

Reviewer bidding behavior was simulated by first sampling the number of bids per reviewer from a normal distribution. Half of these bids were allocated to papers within the reviewer’s own topical area. Bid values were drawn from distributions that favor positive scores for same-area papers and are uniform across other topical areas. Following the convention of \cite{AAAI2021}, we incorporate bidding information by exponentiating the base aggregated affinity score using a bid-dependent factor:
\[
\text{aggscore} = (\text{base aggregated score})^{\text{bidscore}} .
\]
We assign bid scores of $20$, $1$, $0.67$, $0.4$, and $0.25$ to the bid categories \emph{not willing}, \emph{not entered}, \emph{in a pinch}, \emph{willing}, and \emph{eager}, respectively. This transformation preserves scores within $[0,1]$: exponents greater than $1$ penalize assignments (e.g., \emph{not willing}), while exponents less than $1$ amplify assignments (e.g., \emph{in a pinch}, \emph{willing}, \emph{eager}).

Additional reviewer metadata, including seniority and geographic region, were sampled independently at random. Authorship information---including the number of authored papers and coauthors---was generated using Poisson distributions, yielding realistic collaboration structures. Conflicts of interest were introduced by sampling a reviewer-specific number of conflicting papers from an exponential distribution, capturing the heavy-tailed connectivity patterns commonly observed in real reviewer--author networks.


\section{Experiment on Other Datasets} \label{app:other_datasets}

Table \ref{tab:ICLR} shows the results of different algorithms on the ICLR dataset, which has $911$ papers and $2{,}435$ reviewers. Table \ref{tab:s2orc} shows the results on the S2ORC dataset, which has $2{,}446$ papers and $2{,}483$ reviewers. Results on the ICLR and S2ORC datasets largely mirror those observed on the large synthetic dataset. RAMP incurs slightly higher runtime than Default and PLRA, but remains substantially faster than MILP and PM, while outperforming competing methods on most diversity-related metrics at a modest cost in assignment quality.



\begin{table*}[htbp]
\centering
\small
\setlength{\tabcolsep}{6pt}
\renewcommand{\arraystretch}{1.1}

\begin{tabular}{lcccc|c}
\toprule
Metric 
& Default 
& MILP 
& PM 
& PLRA 
& RAMP \\
\midrule
Runtime (s) & 26.006 & 66.945 & 81.846 & 27.303 & 34.115 \\
Quality     & 1.0 & 0.993 & 0.988 & 0.995 & 0.982 \\
Support     & 3644 & 3644 & 12956 & 4585 & 12041 \\
Entropy     & 0.000 & 0.000 & 3339.650 & 692.393 & 3381.906 \\
Diversity   & 0.735 & 0.900 & 0.727 & 0.731 & 0.918 \\
Coauthors   & 19 & 4 & 21 & 22 & 6 \\
2Cycles     & 1 & 0 & 1 & 2 & 0 \\
\bottomrule
\end{tabular}

\caption{Comparison of algorithms on the ICLR dataset}
\label{tab:ICLR}
\end{table*}

\begin{table*}[htbp]
\centering
\small
\setlength{\tabcolsep}{6pt}
\renewcommand{\arraystretch}{1.1}

\begin{tabular}{lcccc|c}
\toprule
Metric & Default & MILP & PM & PLRA & RAMP \\
\midrule
Runtime (s) & 77.031 & 174.841 & 221.858 & 74.435 & 128.582 \\
Quality           & 1.0 & 0.980 & 0.978 & 0.997 & 0.969 \\
Support size      & 9784 & 9784 & 49113 & 12676 & 43645 \\
Entropy           & 0.000 & 0.000 & 12643.684 & 2009.879 & 12901.315 \\
Diversity         & 0.733 & 0.990 & 0.741 & 0.737 & 0.951 \\
Coauthors         & 329 & 8 & 242 & 323 & 19 \\
2Cycles           & 170 & 12 & 150 & 162 & 1 \\
\bottomrule
\end{tabular}

\caption{Comparison of algorithms on the S2ORC dataset}
\label{tab:s2orc}
\end{table*}





\begin{table*}[htbp]
\centering
\small
\setlength{\tabcolsep}{6pt}
\renewcommand{\arraystretch}{1.15}

\begin{tabular}{lcccc|c}
\toprule
Metric & Default & MILP & PM & PRLA & RAMP \\
\midrule
Runtime (s)   & 3.096 & 4.720 & 8.773 & 3.136 & 4.530 \\
Quality       & 1.0 & 0.991 & 0.979 & 0.992 & 0.965 \\
Support size  & 2104 & 2104 & 5609 & 2655 & 5650 \\
Entropy       & 0.000 & 0.000 & 1422.213 & 403.450 & 1628.143 \\
Diversity     & 0.746 & 0.872 & 0.743 & 0.752 & 0.930 \\
Coauthors     & 18 & 0 & 29 & 21 & 3 \\
2Cycles       & 13 & 7 & 11 & 13 & 1 \\
\bottomrule
\end{tabular}

\caption{Performance comparison of algorithms on the AAMAS dataset}
\label{tab:aamas}
\end{table*}

\section{Handling Seniority Constraints}
\label{sec:appendix-sen}

A natural question is how to enforce seniority requirements in reviewer assignment. Earlier, we proposed handling seniority in the same manner as geographic diversity, namely through soft constraints. An alternative approach is a two-stage algorithm, in which senior reviewers are assigned first, followed by junior reviewers. This strategy is widely used in practice by many conferences and has been discussed previously in \cite{AAAI2021}.

From \Cref{tab:reward-seniority-comparison}, we observe that although the soft-constraint approach achieves marginally better solution quality and runtime, it is less effective at satisfying seniority requirements. Modern large-scale conferences, however, may impose strict seniority constraints, preferring to assign at least one senior reviewer to each paper even at the cost of slightly reduced quality or increased runtime.

\begin{table}[H]
\centering
\setlength{\tabcolsep}{5pt}
\renewcommand{\arraystretch}{1.05}
\begin{tabular}{lcc}
\toprule
\textbf{Metric} & \textbf{$\lambda_{\text{sen}} = 0.2$} & \textbf{Two-stage} \\
\midrule
Running time (s) & 122.733 & 142.080 \\
Quality & 1.000 & 0.998 \\
Support size & 43{,}645 & 42{,}946 \\
Entropy & 12{,}901.315 & 12{,}694.372 \\
Diversity & 0.949 & 0.971 \\
Seniority & 0.815 & 1.000 \\
Coauthors & 23 & 28 \\
2Cycles & 1 & 0 \\
\bottomrule
\end{tabular}
\caption{Comparison of single-stage and two-stage methods for handling seniority requirements on the S2ORC dataset}
\label{tab:reward-seniority-comparison}
\end{table}

\section{Hyperparameter Tuning}
\label{sec:hyperparam}

\Cref{tab:hyperparam-comparison-full} presents the full version of \Cref{tab:hyperparam-comparison}. We first set all hyperparameters to zero, and tune the parameters one at the time. Note that quality is relative to the default algorithm on the same dataset

\begin{table*}[htbp]
\centering
\small
\setlength{\tabcolsep}{4pt}
\renewcommand{\arraystretch}{1.05}
\begin{tabular}{lcccc}
\toprule
Metric
& $\text{all} = 0.0$
& $\lambda_{\text{div}} = 0.2$
& $\lambda_{\text{co}} = 0.2$
& $\lambda_{\text{cyc}} = 0.15$ \\
\midrule
Runtime (s) & 113.047 & 124.968 & 113.043 & 112.330 \\
Quality & 0.977 & 0.971 & 0.976 & 0.975\\
Support size & 41{,}231 & 43{,}373 & 41{,}392 & 41{,}351 \\
Entropy & 12{,}308.055 & 12{,}863.558 & 12{,}362.858 & 12{,}292.501 \\
Diversity & 0.835 & 0.949 & 0.832 & 0.846 \\
Seniority & 0.822 & 0.816 & 0.825 & 0.822 \\
Coauthors & 76 & 55 & 30 & 80 \\
2Cycles & 101 & 88 & 104 & 0 \\
\bottomrule
\end{tabular}
\caption{Hyperparameter tuning on the S2ORC dataset}
\label{tab:hyperparam-comparison-full}
\end{table*}

As discussed earlier, the tradeoffs between quality and individual hyperparameters are illustrated in Figures~\ref{fig:2cycle}, \ref{fig:coauth}, and \ref{fig:div}. Normalized quality refers to the quality achieved by the algorithm relative to the maximum attainable quality, which is obtained using the Default algorithm. The fluctuations observed in these curves arise from the randomized nature of the algorithm, which produces different matches between runs.


\begin{figure}[H]
    \centering
    \includegraphics[width=0.75\linewidth]{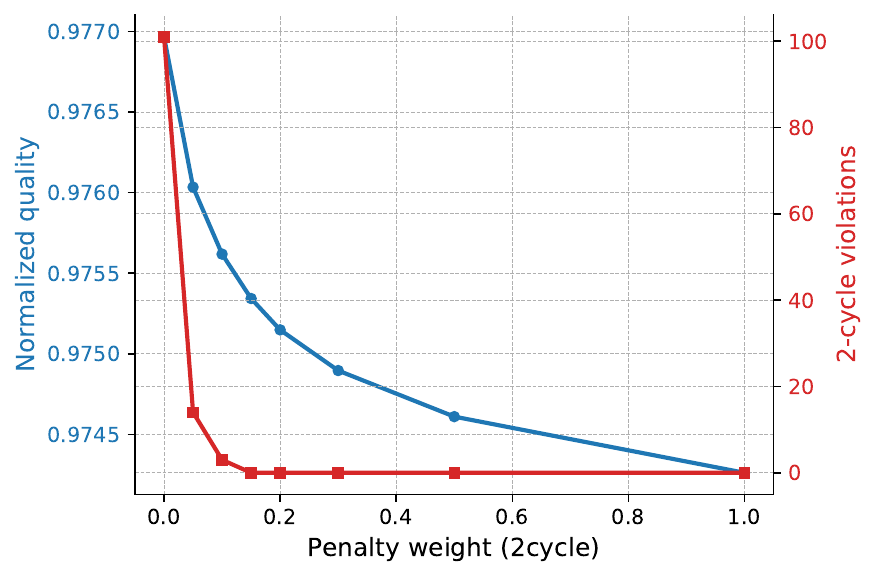}
    \caption{Increasing $\pencycle$ reduces solution quality while simultaneously decreasing the number of 2-cycle violations}
    \label{fig:2cycle}
\end{figure}


\begin{figure}[H]
    \centering
    \includegraphics[width=0.75\linewidth]{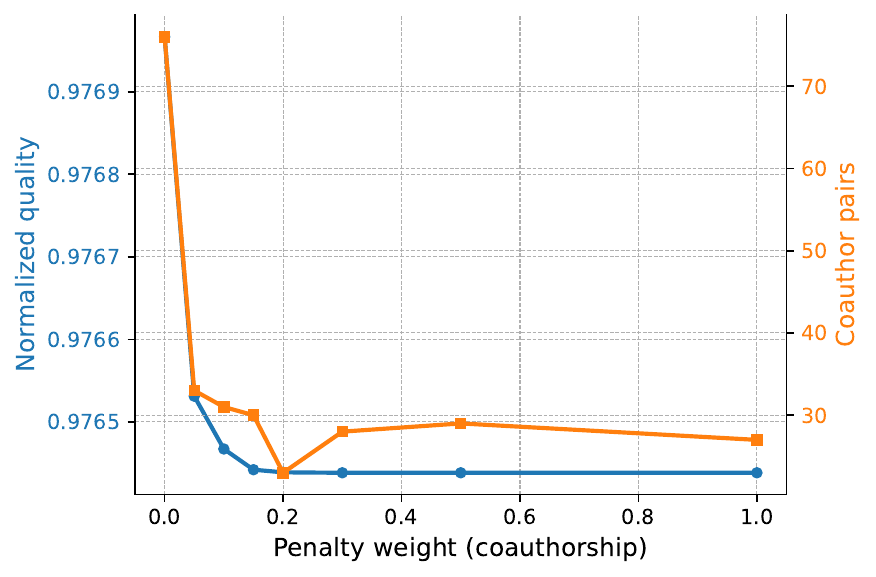}
    \caption{Increasing $\pencoauthor$ reduces solution quality and the number of coauthor-pair violations}
    \label{fig:coauth}
\end{figure}


\begin{figure}[H]
    \centering
    \includegraphics[width=0.75\linewidth]{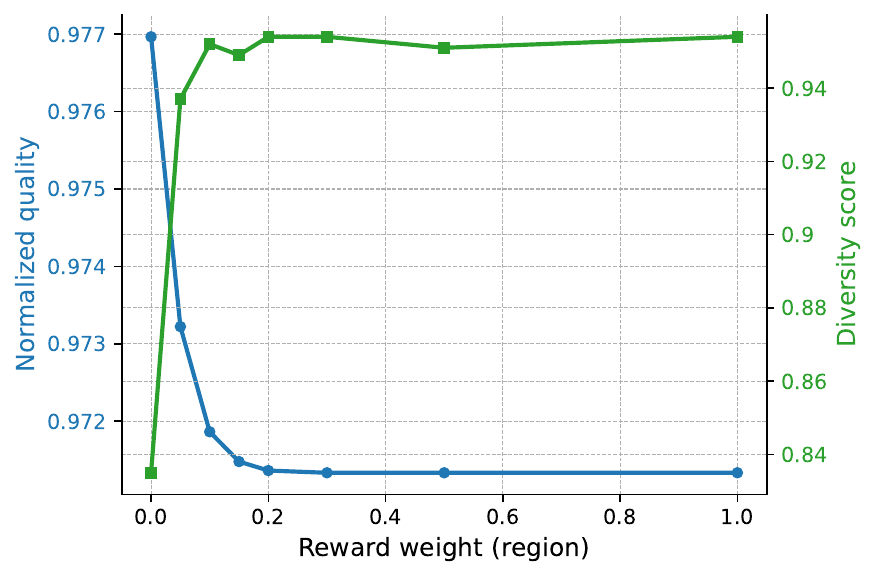}
    \caption{Increasing $\rewarddiv$ rapidly improves diversity, with diminishing returns beyond approximately $0.2$}

        \label{fig:div}
\end{figure}

In addition to the parameters above, the hyperparameter $Q$ can also be tuned. Results are plotted in Figure \ref{fig:q} Experiments on the S2ORC dataset indicate that both runtime and solution quality increase as $Q$ grows; however, these effects are negligible in magnitude. In practice, program chairs are therefore more likely to tune $Q$ to mitigate collusion concerns rather than to optimize runtime or quality. The choice of perturbation function is similarly tunable and was explored extensively in \cite{PM}.

\begin{figure}[H]
        \centering
        \includegraphics[width=0.6\linewidth]{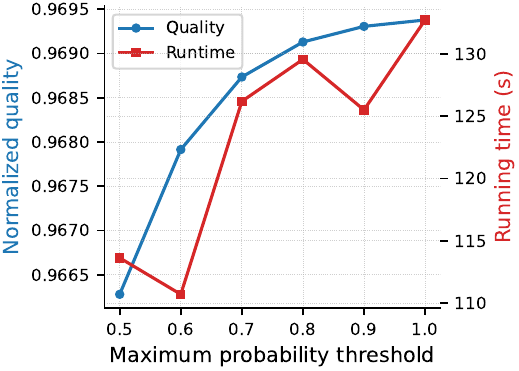}
        \caption{Quality and runtime as a function of $Q$ on the S2ORC dataset}
        \label{fig:q}
\end{figure}



\section{Extended Ablation Results}
\label{app:ablation}

This section reports the complete results for the ablation studies summarized in Section~\ref{subsec:ablation_piecewise_linear_and_sampling}, including piecewise linearization and attribute-aware sampling.

\begin{table}[htbp]
\centering
\begin{tabular}{lrr} 
\toprule
Metric & PWL & Non-PWL \\
\midrule
Runtime (s)   & 1088.9 & 16569.9 \\
Quality       & 1.0 & 0.999 \\
Support size  & 537266 & 825421 \\
Entropy       & 143587 & 158028 \\
Diversity     & 0.896 & 0.893 \\
Coauthors & 26 & 27 \\
2Cycles    & 0 & 2 \\
\bottomrule
\end{tabular}
\caption{Full results for piecewise linearization (extension of \Cref{tab:pwl_ablation})}
\label{tab:pwl_ablation_full}
\end{table}

\begin{table}[htbp]
\centering
\begin{tabular}{lrr}
\toprule
Metric & Attribute-Aware & Vanilla \\
\midrule
Runtime (s)   & 1088.9 & 1083.2 \\
Quality       & 1.0 & 1.0 \\
Support size  & 537266 & 537266 \\
Entropy       & 143587 & 143587 \\
Diversity     & 0.896 & 0.697 \\
Coauthors & 26 & 199 \\
2Cycles    & 0 & 0 \\
\bottomrule
\end{tabular}
\caption{Full results for attribute-aware sampling (extension of \Cref{tab:sampling_ablation})}
\label{tab:sampling_comparison_full}
\end{table}

\section{Attribute-Aware Sampling Algorithms}
\label{app:sampling_algo}

In this section, we provide the detailed pseudocode for the Attribute-Aware Sampling procedure described in Section~\ref{subsec:sampling}. This procedure is built upon the sampling framework of Jecmen et al.~\cite{PLRA}. 

The core mechanism involves iteratively decomposing the fractional assignment matrix into paths and cycles to shift probability mass until an integral solution is reached. Our key modification to the original algorithm lies in the search subroutine: whereas the standard approach selects paths and cycles arbitrarily, our method utilizes an attribute-aware search (Algorithm~\ref{alg:find_chain}). This subroutine prioritizes linking ``similar'' reviewers (e.g., past co-authors or those from the same region) within the same update step. By coupling the rounding decisions for these reviewers, the algorithm minimizes the variance that leads to violations of soft diversity and co-authorship constraints.

Algorithm~\ref{alg:sampling} presents the main iterative rounding loop, and Algorithm~\ref{alg:find_chain} details the priority-based Depth-First Search (DFS) strategy.

\begin{algorithm}[h]
\caption{Attribute-Aware Sampling (Path or Cycle)}
\label{alg:sampling}
\begin{algorithmic}[1]
\STATE \textbf{Input:} Fractional assignment matrix $\mathbf{x} \in [0, 1]^{|\mathcal{P}| \times |\mathcal{R}|}$.
\STATE \textbf{Output:} Integral assignment matrix $\mathbf{X} \in \{0, 1\}^{|\mathcal{P}| \times |\mathcal{R}|}$.
\WHILE{exists $(p, r)$ such that $0 < x_{p,r} < 1$}
    \STATE Construct the bipartite graph $G = (\mathcal{P} \cup \mathcal{R}, E)$ where $E = \{(p, r) \mid 0 < x_{p,r} < 1\}$
    \STATE $C \leftarrow \text{FindAttributeAwareChain}(G, \mathbf{x})$ (See \Cref{alg:find_chain})
    \STATE Partition edges of $C$ into $C_{odd}$ and $C_{even}$ based on their position in the sequence.
    \STATE Calculate maximum feasible steps:
    \STATE $\quad \alpha \leftarrow \min \big( \{1 - x_{e} \mid e \in C_{odd}\} $
    \STATE $\qquad \qquad \cup \{x_{e} \mid e \in C_{even}\} \big)$
    \STATE $\quad \beta \leftarrow \min \big( \{x_{e} \mid e \in C_{odd}\} $
    \STATE $\qquad \qquad \cup \{1 - x_{e} \mid e \in C_{even}\} \big)$
    
    \IF{$C$ is a Path starting at $u$ and ending at $v$}
        \STATE Update $\alpha, \beta$ ensuring reviewer load constraints at $u, v$ are not violated.
    \ENDIF
    
    \STATE With probability $\frac{\beta}{\alpha + \beta}$:
        \STATE \quad $x_{e} \leftarrow x_{e} + \alpha$ for $e \in C_{odd}$
        \STATE \quad $x_{e} \leftarrow x_{e} - \alpha$ for $e \in C_{even}$
    \STATE Else (with probability $\frac{\alpha}{\alpha + \beta}$):
        \STATE \quad $x_{e} \leftarrow x_{e} - \beta$ for $e \in C_{odd}$
        \STATE \quad $x_{e} \leftarrow x_{e} + \beta$ for $e \in C_{even}$
\ENDWHILE
\RETURN $\mathbf{x}$
\end{algorithmic}
\end{algorithm}

\begin{algorithm}[h]
\caption{FindAttributeAwareChain (Subroutine)}
\label{alg:find_chain}
\begin{algorithmic}[1]
\STATE \textbf{Input:} Bipartite graph $G$, Assignment $\mathbf{x}$.
\STATE \textbf{Output:} A sequence of edges forming a Cycle or Path.
\STATE $\text{visited} \leftarrow \emptyset$, $\text{stack} \leftarrow []$.
\STATE \textbf{Selection of Start Node:}
\STATE \quad If there exists a node $v$ with odd degree in $G$ (fractional load), let $v_{start} = v$.
\STATE \quad Else, pick arbitrary $v_{start}$ with degree $>0$.
\STATE Push $v_{start}$ to $\text{stack}$.

\WHILE{$\text{stack}$ is not empty}
    \STATE $u \leftarrow \text{stack.peek()}$
    \STATE $V_{\text{adj}} \leftarrow \{v \mid (u, v) \in E\} \setminus \{\text{parent}(u)\}$
    
    \IF{$V_{\text{adj}} = \emptyset$}
        \STATE \COMMENT{Path found}
        \RETURN $\text{stack}$ as a Path.
    \ENDIF

    \IF{$u \in \mathcal{P}$}
        \STATE \COMMENT{\textbf{Attribute-Awareness:} Prioritize similar reviewers}
        \STATE Let $r_{prev}$ be the reviewer preceding $u$ in $\text{stack}$.
        \STATE Construct ordered list $L$ from $V_{\text{adj}}$ based on priority:
        \STATE \quad 1. $v \in \mathcal{N}(r_{prev})$ \hfill (Co-authors)
        \STATE \quad 2. $\text{Region}(v) = \text{Region}(r_{prev})$ \hfill (Same Region)
        \STATE \quad 3. Others
    \ELSE
        \STATE \COMMENT{Reviewer node: standard traversal}
        \STATE Construct list $L$ from $V_{\text{adj}}$
    \ENDIF
    
    \STATE Pick first $v$ in $L$.
    \IF{$v \in \text{visited}$}
        \STATE \COMMENT{Cycle found}
        \STATE Extract cycle portion from $\text{stack}$ (from first occurrence of $v$ to $u$).
        \RETURN Cycle.
    \ELSE
        \STATE Push $v$ to $\text{stack}$, mark $v$ visited.
    \ENDIF
\ENDWHILE
\end{algorithmic}
\end{algorithm}

\end{document}